\documentclass[12pt,english,floatfix,superscriptaddress,aps,preprint,showpacs]{revtex4}
\usepackage{amsmath}
\usepackage{amssymb}
\usepackage{amsbsy}
\usepackage{amsfonts}
\usepackage{amsopn}
\usepackage{amstext}
\usepackage{graphicx}
\usepackage{amssymb}
\usepackage{amsfonts}
\usepackage{amsmath}
\usepackage{graphicx}
\usepackage[english]{babel}
\usepackage{color}
\usepackage[dvips]{epsfig}
\usepackage[dvips]{graphicx}
\usepackage{float}
\usepackage{units}
\usepackage{textcomp}
\usepackage{babel}

\begin{document}

\title{Graviton resonances on two-field thick branes}
\author{W. T. Cruz}
\email{wilamicruz@gmail.com}
\affiliation{Instituto Federal de Educa\c{c}\~{a}o, Ci\^{e}ncia e Tecnologia do Cear\'{a} (IFCE), Campus Juazeiro do Norte - 63040-540 Juazeiro do Norte-Cear\'{a}-Brazil}
\author{L. J. S. Sousa}
\email{luisjose@fisica.ufc.br}
\affiliation{Instituto Federal de Educa\c{c}\~{a}o, Ci\^{e}ncia e Tecnologia do Cear\'{a} (IFCE),
Campus Canind\'{e}, Canind\'{e}-Cear\'{a}-Brazil}
\author{R. V. Maluf}
\email{r.v.maluf@fisica.ufc.br}
\affiliation{Departamento de F\'{i}sica - Universidade Federal do Cear\'{a} (UFC) - C.P. 6030, 60455-760
Fortaleza-Cear\'{a}-Brazil}
\author{C. A. S. Almeida}
\email{carlos@fisica.ufc.br}
\affiliation{Departamento de F\'{i}sica - Universidade Federal do Cear\'{a} (UFC) - C.P. 6030, 60455-760
Fortaleza-Cear\'{a}-Brazil}
\date{\today}

\begin{abstract}
This work presents new results about the graviton massive spectrum in two-field thick branes. Analyzing the massive spectra with a relative probability method we have firstly showed the presence of resonance structures and obtained a connection  between the thickness of the defect and the lifetimes of such resonances. We obtain another interesting results considering the degenerate Bloch brane solutions. In these thick brane models, we have the emergence of a splitting effect controlled by a degeneracy parameter.  When the degeneracy constant tends to a critical value, we have found massive resonances to the gravitational field indicating the existence of modes highly coupled to the brane. We also discussed the influence of the brane splitting effect over the resonance lifetimes.
\end{abstract}


\maketitle

\section{Introduction\label{intro}}

Recently, it has been given much attention to the study of topological defects in the context of brane-world models, due to its property of allowing the localization of several different types of fields. As extended defects in field theory the domain walls have been used in high-energy physics to represent brane scenarios with extra dimensions \cite{rub,vis}. The Bloch walls, which could be seen as chiral interfaces \cite{wall}, are used in the context of extra dimensions to construct a $(4,1)D$ model of two scalar fields coupled with gravity, the so called Bloch brane \cite{dionisio}.

The Bloch brane model is generated dynamically and has internal structure. The asymptotic bulk metric is a slice of a five-dimensional anti-de Sitter (AdS) spacetime, denoted by $AdS_5$. Such scenario may be used to mimic a brane-world containing internal structure \cite{aplications}, which have implications on the density of matter-energy along the extra dimension \cite{brane}. The appearance of the internal structure could be also observed by a splitting effect on the curvature invariant.

On the other hand, Dutra et al. have showed that the Bloch brane scenario addressed in \cite{dionisio}, holds more general soliton solutions \cite{degen1, degen2, degen3}. The brane configurations obtained from these new solutions were named by the authors as \textit{degenerate Bloch branes} due to the existence of a degeneracy parameter that is not present in the Lagrangian density. Such defects present more details as the appearing of two-kink solutions. These additional features were interpreted as the formation of a double wall structure. When the degeneracy parameter approaches to a critical value, the brane splits in two and its separation become larger. This effect will contribute to the emergence of massive graviton resonances. The phenomenon of two separate interfaces on the defect is known in condensed matter physics as complete wetting \cite{wetting1, wetting2}. Moreover, the same behavior was considered as a critical phenomena of phase transition on thick branes in warped geometries \cite{fase}.

In the above scenarios some authors have investigated the localization of several types of bulk fields. Namely, fermion fields \cite{degen3,castro,ca,chineses0}, gauge fields \cite{nosso6, nosso7} and graviton zero mode \cite{dionisio, degen2}. However, the study of the graviton massive spectrum has not been properly studied. It is worthwhile to mention that,  the search for resonances in warped spacetimes \cite{ca,nosso6,gremm,nosso,csaki1,chineses1,chineses2,chineses3,chineses4,chineses5,csaki2} has received attention because they give us important information about the interaction of Kaluza-Klein (KK) massive spectrum with the four-dimensional brane. Specifically in the study of gravity localization, the presence of a resonance at zero energy is related to the existence of a large-distance region on which the 4D laws of gravity are valid \cite{csaki2, metastable}. Therefore, if the resonance width becomes very large it results in nonphysical effects.

To the best of our knowledge the first work about localization of graviton zero mode on the Bloch brane was the work of Bazeia and Gomes \cite{dionisio}. Their results were also confirmed on the degenerate Bloch branes by Dutra et al. \cite{degen2}. However, the important issues concerning the analysis of the massive spectrum and the search for resonances in these scenarios have not been addressed. Therefore, in the present work we propose to study the graviton massive spectrum and search for resonant states on the two field thick brane scenarios described above. Additionally, we intend to verify the relation between the brane thickness and the internal structure over possible detected resonances.

We organized this work as follows. In Sec. \ref{sec.bloch}, we review the Bloch brane scenario and search for new resonant structures in this model in Sec. \ref{sec.massive}. We addressed the degenerate Bloch brane solutions in the beginning of the Sec. \ref{sec.other} and we search for new graviton resonances in this more general setup. Finally, we present our results and conclusions in Sec. \ref{sec.conc}.

\section{Brane setup\label{sec.bloch}}
The two-field thick brane scenario that we consider is composed by two
fields $\phi$ and $\chi$ coupled to gravity which depend only on the extra dimension
$y$. Such model was previously studied in references \cite{dionisio,castro,ca,nosso6}. Their action is given as follows
\begin{equation}\label{eq:action}
S=\int d^{5}x\sqrt{-G}\Big[-\frac{1}{4}R+\frac{1}{2}(\partial\phi)^{2}+\frac{1}{2}(\partial\chi)^{2}-V(\phi,\chi)\Bigr],
\end{equation} where $R$ is the scalar curvature and the spacetime is an AdS $D=5$ with metric
\begin{equation}\label{eq:metric}
ds^{2}=e^{2A(y)}\eta_{\mu\nu}dx^{\mu}dx^{\nu}-dy^{2}.
\end{equation}
The Minkowski spacetime metric is $\eta_{\mu\nu}$ with signature $(-,+,+,+)$ and the indices $\mu$, $\nu$ run over $1$ to $4$.

The corresponding equations of motion are
\begin{eqnarray}
\phi'^{2}+\chi'^{2}-2V(\phi,\chi) & = & 6A'^{2}\label{eq:eqmov1}\nonumber\\
\phi'^{2}+\chi'^{2}+2V(\phi,\chi) & = & -6A'^{2}-3A''\label{eq:eqmov2}\nonumber\\
\xi''+4A^{\prime}\xi' & = & \partial_{\xi}V,\,\,\,\,\,\xi=\phi,\chi,\label{eq:eqmov3}
\end{eqnarray}
where prime stands for derivative with respect to $y$.

\begin{figure}[!htb] 
       \begin{minipage}[b]{0.48 \linewidth}
           \includegraphics[width=\linewidth]{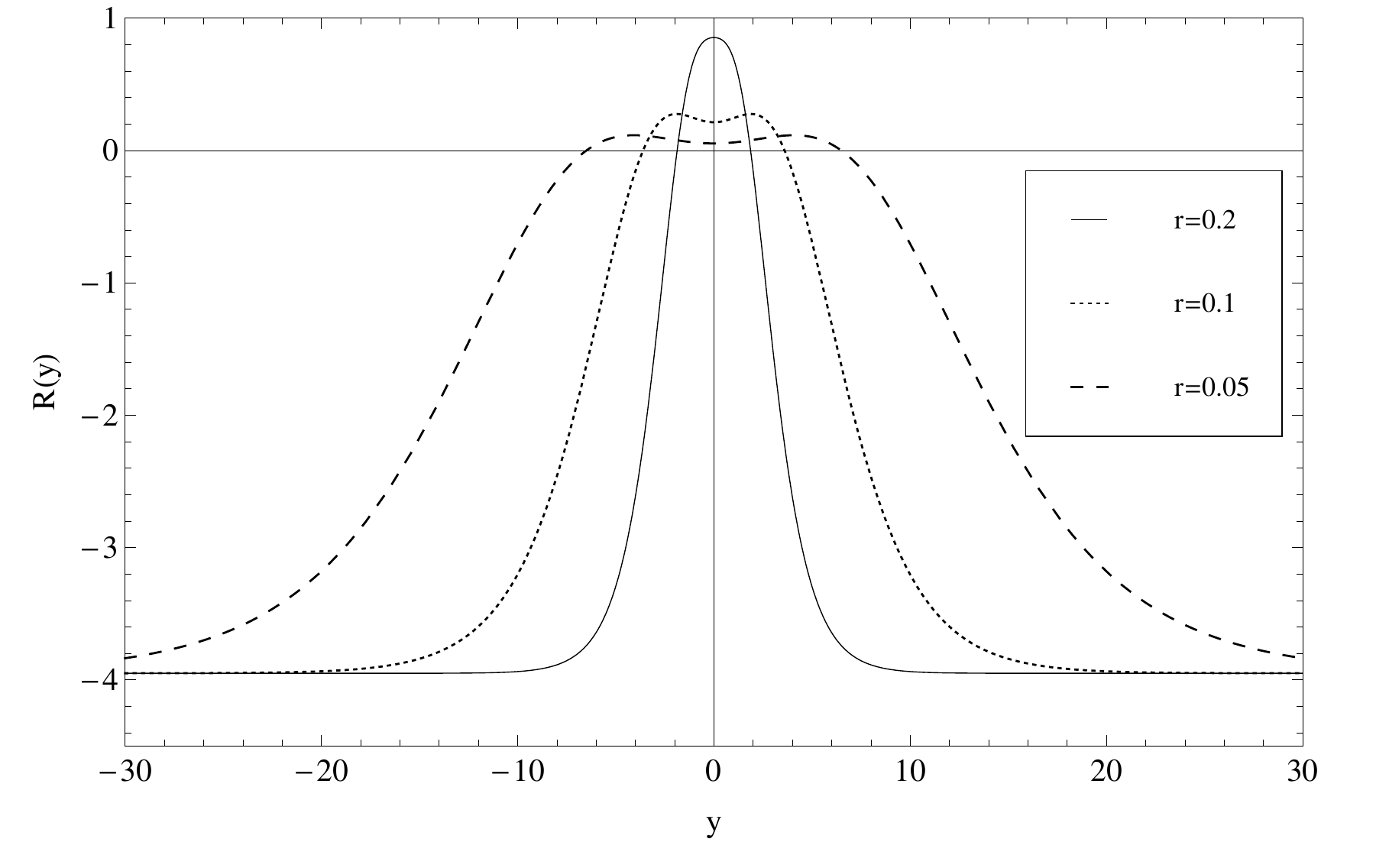}\\
           \caption{Plots of the curvature invariant $R(y)$ with $r=0.2$, $0.1$ and $0.05$.}
          \label{fig:gravbloch0}
       \end{minipage}\hfill
       \begin{minipage}[b]{0.48 \linewidth}
           \includegraphics[width=\linewidth]{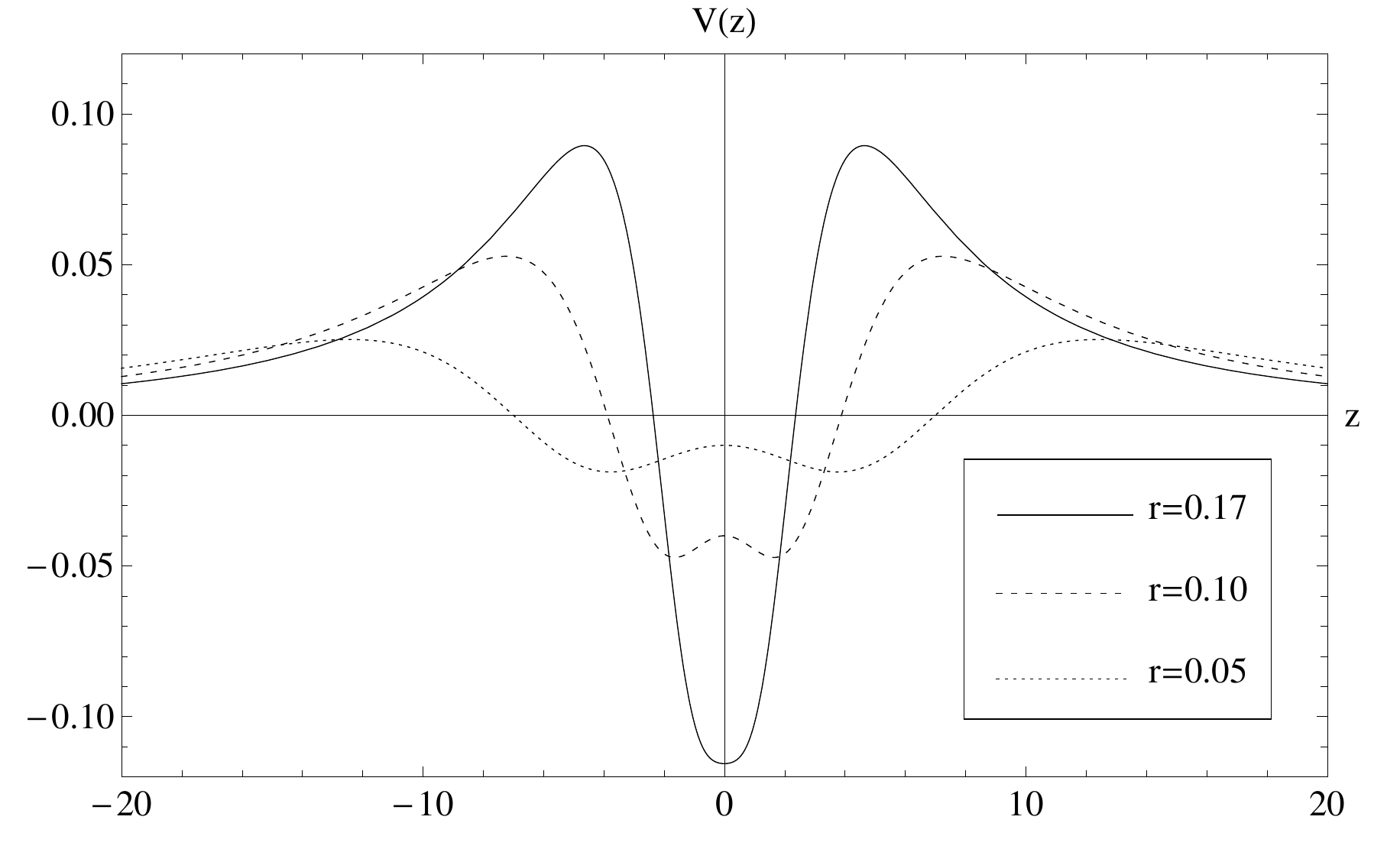}\\
           \caption{Plots the potential $V(z)$ with $r=0.17,0.10,0.05$. We note the appearance two separate minima at z=0, indicating the presence of internal structure.}
           \label{fig:gravblochpot}
       \end{minipage}
   \end{figure}

A method for solving the coupled differential equations system \eqref{eq:eqmov3} has been developed in the context of thick branes \cite{bazeia1,bazeia2,bazeia3,bazeia4,shif,alonso}. It consists of a appropriate redefinition of the potential $V(\phi,\xi)$ as
\begin{equation} V(\phi,\chi)=\frac{1}{8}\left[\left(\frac{\partial W}{\partial\phi}\right)^{2}+\left(\frac{\partial W}{\partial\chi}\right)^{2}\right]-\frac{1}{3}W^{2}
\end{equation}
in terms of a superpotential
\begin{equation}\label{eq:supot}
W(\phi,\chi)=2\phi-\frac{2}{3}\phi^{3}-2r\phi\chi^{2}.
\end{equation}
This implies that the resulting first-order equations can be written as  $\phi^{\prime}=\frac{1}{2}\frac{\partial W}{\partial\phi}$,
$\chi^{\prime}=\frac{1}{2}\frac{\partial W}{\partial\chi}$ and $A^{\prime}=-\frac{1}{3}W$, from which we find the solutions that describe our brane model, namely
\begin{equation}
\phi(y)=\tanh(2ry),\label{phi}
\end{equation}
\begin{equation}
\chi(y)=\sqrt{\left(\frac{1}{r}-2\right)}{\rm sech}(2ry)\label{chi},
\end{equation}
and
\begin{equation}
A(y)=\frac{1}{9r}\Bigl[(1-3r)\tanh^{2}(2ry)-2\ln\cosh(2ry)\Bigr].\label{warp}
\end{equation}
From Eq. \eqref{chi}, we can see that for the limit $r=0.5$ the one-field scenario is recovered. For certain values of the coupling parameter $r$ that controls the brane thickness, we have a splitting of the defect and the appearance of an internal structure. This characteristic is also evident on the curvature invariant. For this geometry, for instance, we obtain
\begin{equation}
R=-\left[8A''+20(A')^2\right].
\end{equation}
The Ricci scalar is finite, which we can observe through the Fig. (\ref{fig:gravbloch0}). For $0.5>r>r_c$, with $r_c \approx 0.17 $ there is a maximum at $y=0$. When $r$ arrives to the interval $r_c>r>0$, the maximum splits into two separate maxima, which indicates the presence of the internal structure. These effects will influence the behavior of the massive modes, which will be investigated in the following section.

\section{Massive spectrum and resonances}\label{sec.massive}

The seminal work of Bazeia and Gomes \cite{dionisio} considering the Bloch brane model performed an analysis in the zero modes of the graviton but have not confirmed the existence of resonant modes. From now, we will take this scenario again focusing our attention in the massive spectrum and seeking resonant modes for the graviton.
In order to search for resonances in the massive spectrum we must obtain a Shr\"{o}edinger-like equation to the graviton on the fifth dimension. Initially, we perform a metric perturbation using
$ ds^2=e^{2A(y)}(\eta_{\mu\nu}+\epsilon h_{\mu\nu})dx^\mu
dx^\nu-dy^2$, where $h_{\mu\nu}=h_{\mu\nu}(x,y)$ represents the graviton with the axial gauge $h_{5N}=0$. When we set the metric fluctuation as transverse and traceless (TT), namely $\overline{h}_{\mu\nu}$, its equations of motion take the simplified form \cite{gremm, kehagias, de}:
\begin{equation}\label{eq:grav}
\overline{h}_{\mu\nu}^{\prime\prime}+4A^{\prime}
\overline{h}_{\mu\nu}^{\prime}=e^{-2A}\partial^2\overline{h}_{\mu\nu}^{\prime},
\end{equation}
where $\partial^2$  is the four-dimensional wave operator. Using the transformation $dz=e^{-A(y)}dy$ and choosing an ansatz containing a bulk wave function times a space plane wave, $\overline{h}_{\mu\nu}(x,z)=e^{ip\cdot
x}e^{-\frac{3}{2}A(z)}\psi_{\mu\nu}(z)$, we can rewrite the Eq. \eqref{eq:grav} as a
Schr\"{o}dinger-like equation given by
\begin{equation}\label{eq:schro}
-\frac{d^2\psi(z)}{dz^2}+V(z)\,\psi(z)=m^2\,\psi(z),
\end{equation}
with the potential $V(z)=\frac32\,A^{\prime\prime}(z)+\frac94\,A^{\prime2}(z)$. The Eq. (\ref{eq:schro}) leads to no tachyonic states and owns a normalizable zero mode solution as was showed in Ref. \cite{dionisio}.

When $m^2\gg V_{max}$, the potential  represents only a small perturbation and the solutions will acquire plane wave structure. In Fig. \ref{fig:gravblochpot} we plot the potential $V(z)$ varying $r$. As in the curvature scalar (Fig. \ref{fig:gravbloch0}) we note the appearance of  a splitting effect due to the thickness of the defect.  The minimum of the potential separates in two as we reduce $r$, acquiring the shape of a double-well-type potential. Similar feature was also found in the potential of the Shr\"{o}edinger equation for the TT sector
of the metric perturbations in a thick brane scenario generated by one scalar field \cite{fase}.

For some specific energies, the solutions of Eq. \eqref{eq:schro} could exhibit large amplitudes inside the brane in comparison with its values far from the defect. These solutions are interpreted as resonant modes and reveal states with large probabilities to be found on the brane \cite{gremm}.

From the Eq. (\ref{eq:schro}) we can consider $\zeta|\psi_{m}(0)|^2$  as the probability for finding the mode on the brane, where $\zeta$ is a normalization constant. In this way, we must know the variation of this quantity in terms of $m$ in order to find possible resonant modes. For this purpose, we first consider two different methods to detect resonant modes. One of them uses directly the probability of finding the mode on the brane from the square of the normalized wave function \cite{ca,csaki2,nosso1}, namely
\begin{equation}\label{rel}
P(m)=\frac{|\psi_{m}(0)|^2}{\int_{-z_{max}}^{+z_{max}}|\psi_{m}(z)|^2dz}.
\end{equation}

A second method uses the relative probability \cite{chineses1,chineses2,chineses3, chineses4, chineses5}:
\begin{equation}\label{rel_prob}
N(m)=\frac{\int_{-z_{b}}^{+z_{b}}|\psi_{m}(z)|^2}{\int_{-z_{max}}^{+z_{max}}|\psi_{m}(z)|^2dz}.
\end{equation}
In both approaches, the massive modes are considered in a box with borders $|z|=z_{max}$ far from the turning points.

The difference between the two methods  is that in the case of relative probability $N(m)$ we have a narrow integration range around the brane $-z_b<z<z_b$ conveniently chosen such that $z_b=0.1z_{max}$ \cite{chineses3}. The main motivation for using this approach is that, since thick branes are objects spread around the origin of the extra dimension, the probability of finding the modes must be integrated in a small region around the location of the defect in order to detect resonances. Furthermore, according to the formal quantum theory of resonance, we can study the probability of finding the massive modes around the vicinity of the brane position within a relatively large region \cite{ca,chineses1,chineses4}.

\begin{figure}[!htb] 
       \begin{minipage}[b]{0.48 \linewidth}
           \includegraphics[width=\linewidth]{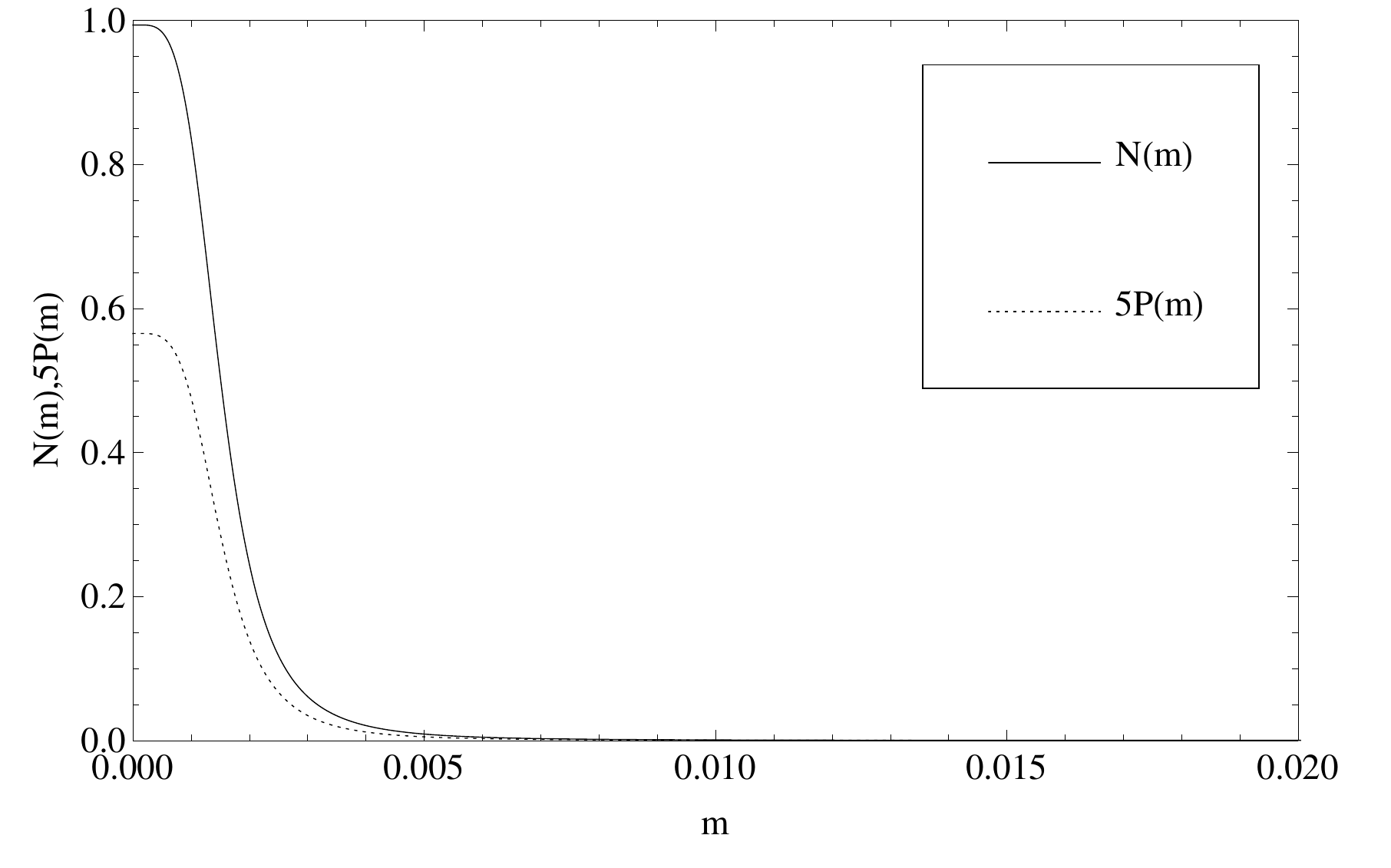}\\
           \caption{Plots of the functions $N(m)$ and $P(m)$ with $r=0.1$. In order to accommodate both graphs in a single frame, the function $P(m)$ appears multiplied by a factor of 5.}
          \label{fig:gravbloch1}
       \end{minipage}\hfill
       \begin{minipage}[b]{0.48 \linewidth}
           \includegraphics[width=\linewidth]{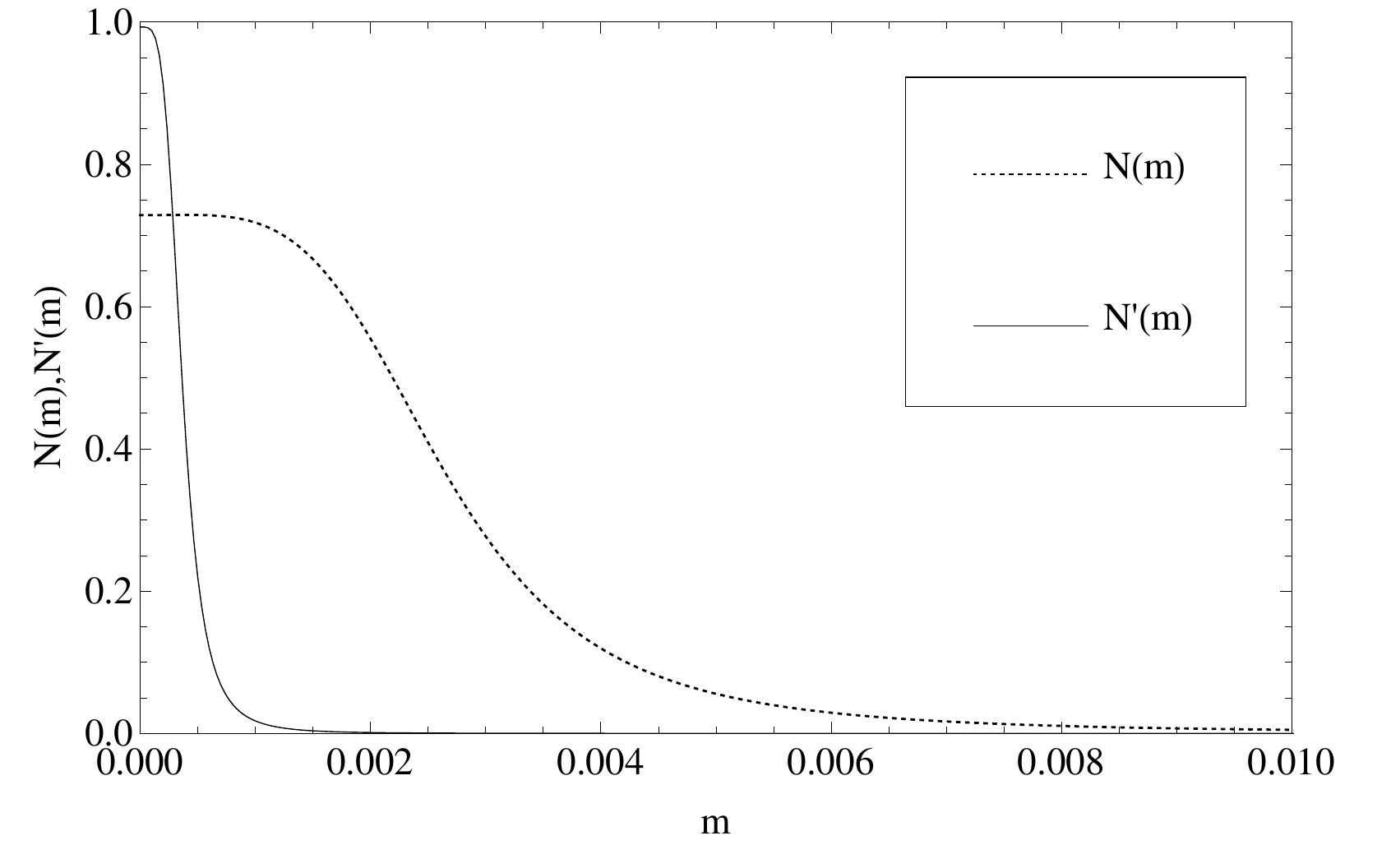}\\
           \caption{Plots of the function $N(m)$ with $z_{max}=200$ and $N'(m)$ with $z_{max}=600$. We have used $r=0.01$.}
           \label{fig:gravbloch2}
       \end{minipage}
   \end{figure}

We must choose one of the probability methods based on their efficiency to detect in a better way the resonant modes on the KK spectrum (if they exist). Such structures are characterized by large amplitudes inside the brane in comparison with the amplitude outside. Thus we first compare the two functions $P(m)$ and $N(m)$ to choose the more suitable. Due to the coordinate transformations and the complexity of the $A(y)$ function, we can not find $A(z)$ analytically. Therefore, we solve the Eq. \eqref{eq:schro} by numerical integration and compare the functions $N(m)$ and $P(m)$. The result is showed in Fig. \ref{fig:gravbloch1}. As expected, the two methods indicate a resonance at $m=0$.

It is worth to mentioning that the two functions $N(m)$ and $P(m)$ acquire different values when $m^2>>V_{max}$ \cite{ca,chineses3}. In this region, where the solutions present plane wave structures, we have $P(m)\rightarrow\frac{1}{z_{max}}$ \cite{ca} and $N(m)\rightarrow \frac{z_b}{z_{max}}$ \cite{chineses3}. We also observed in Fig. \ref{fig:gravbloch1} that the resonance peak is more pronounced and broader on the $N(m)$ function. This characteristic shows that using $N(m)$ we can detect more easily possible narrow resonance structures. Then, we will adopt $N(m)$ in our analysis henceforth.

Since we consider the  KK modes $\psi(z)$ in a box with borders $|z|=z_{max}$, beyond which $\psi(z)$ are turned
into plane waves, the $z_{max}$ value should not interfere with the mass of the resonant modes to maintain the consistency of the method. This is a fundamental characteristic because the physical information is contained in the value of the resonance peak, which should not depends on $z_{max}$, since it is chosen sufficiently large. In Fig. \ref{fig:gravbloch2},  we plot the function $N(m)$ with $z_{max}=200$ and $N'(m)$ with $z_{max}=600$. As expected, the position of the resonant mode in $N(m)$ is not altered when we vary $z_{max}$.

\begin{figure}[!htb] 
      \centering
       \begin{minipage}[b]{1 \linewidth}
           \includegraphics[width=\linewidth]{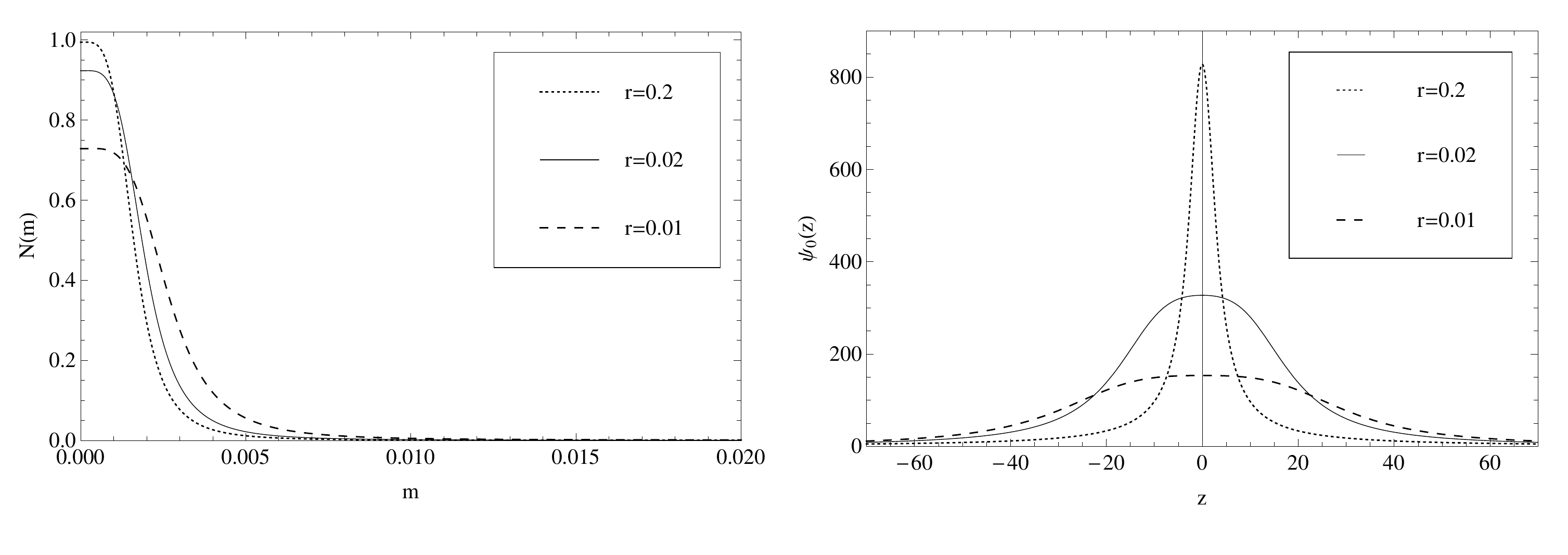}\\
           \caption{Plots of the function $N(m)$ (left) and the zero modes solutions (right) with $r=0.2,0.02$ and $0.01$.}
         \label{fig:gravbloch3}
       \end{minipage}
   \end{figure}

We must now analyze the behavior of the function $N(m)$ as well as its dependence on the thickness of the membrane. The real parameter $r$ controls the thickness of the defect and may give rise to an internal structure on the membrane. This characteristic can be verified by analyzing the structure of the potential as well as the matter energy density in terms of $r$, as shown in Ref. \cite{dionisio}. In order to verify the effects of the brane thickness on the graviton massive modes, we analyze the behavior of the  function $N(m)$ in terms of $r$. The result is shown on the left side of the Fig. \ref{fig:gravbloch3}. We also plot the correspondent resonant modes on right side of the same figure.

By varying the $N(m)$ function in terms of $r$ we find resonances at $m=0$.  We can estimate the lifetime $\tau$ of a resonance by $(\Delta m)^{-1}$, where $\Delta m$ is the width at half maximum in mass of the peaks in $N(m)$ \cite{ca,chineses1,chineses2,chineses3,lifetime}. For small $r$, the resonance becomes very broad characterizing states with lifetimes too short to result in any physical effect. This characteristic can be confirmed observing the $\psi_{0}(z)$ function correspondent to the resonant modes at $m=0$, as can be seen in Fig. \ref{fig:gravbloch3} (right side). In this case, the zero modes are scattered along the extra dimension leading to non-localized zero modes.

\section{Degenerate Bloch branes}\label{sec.other}

In this section we search for graviton resonances in more general thick brane scenarios generated by two scalar fields. As a matter of fact, after the work of Bazeia and Gomes in Bloch branes \cite{dionisio}, the existence of new field solutions coming from the Bloch brane scenario was shown by Dutra and collaborators \cite{degen1, degen2, degen3}.  In these works, the authors directed their analysis to the stability and zero mode solutions of the graviton, but now we generalize the analysis to the massive spectrum in a quantum mechanics scenario. On these new solutions, which were called degenerate Bloch walls (DBW), the brane thickness is controlled by the variation of a parameter associated to the domain wall degeneracy. Another important characteristic of the DBW is the appearance of two-kink solutions indicating the formation of two branes.

Particularly, we choose the solutions presented in Ref. \cite{degen3}, where the action and the metric are the same in Eq. \eqref{eq:action} and Eq. \eqref{eq:metric} respectively. Therefore, the Schr\"{o}dinger-like Eq. \eqref{eq:schro} is also maintained. With the DBW field solutions and the corresponding warp factors, we reevaluated the $N(m)$ function defined in \eqref{rel_prob}, in order to search for resonances on the massive spectrum over these new conditions.

First, we consider the potential $V(\phi ,\chi )$ which is written in terms of a superpotential as
\begin{equation}
V(\phi ,\chi )=\frac{1}{2}\left[ \left( \frac{\partial W}{%
\partial \phi }\right) ^{2}+\left( \frac{\partial W}{\partial
\chi }\right) ^{2}\right] -\frac{4}{3}W^{2},
\end{equation}
which now takes the form
\begin{equation}
\overline{W}(\phi ,\chi )=\phi \left[ \lambda \left( \frac{\phi ^{2}}{3}-a^{2}\right)
+\mu \,\chi ^{2}\right],
\end{equation}
from where we can recover the Eq. \eqref{eq:supot} by choosing $a=1$, $\lambda=-2$ and $\mu=-2r$.

The motivation to consider the superpotential of this form is that now we can control the warp factor behavior with more parameters than the $r$ parameter as in the previous section. This characteristic will bring us richer solutions that will give rise to new resonance structures. For more details on how to obtain the warp factors, the reader should consult the Refs. \cite{degen1, degen2, degen3}.

\begin{figure}[!htb] 
      \centering
       \begin{minipage}[b]{1 \linewidth}
           \includegraphics[width=\linewidth]{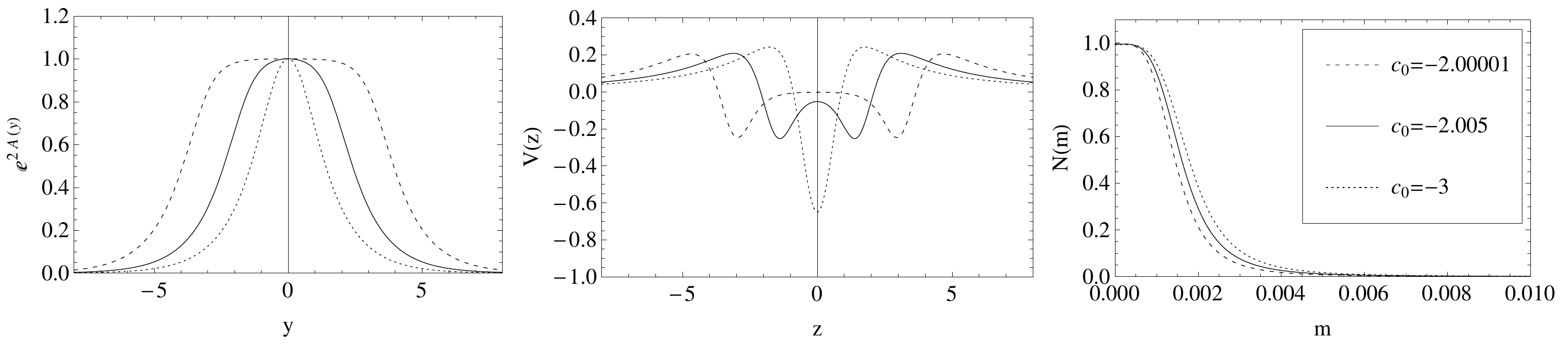}\\
           \caption{Plots of the warp factor (\textit{left}), potential (\textit{center}) and the function $N(m)$ (\textit{right}) for the DBW(1) solution. We have used $a= \mu=1$.}
        \label{fig:resd1}
       \end{minipage}
   \end{figure}

The first case that we will consider, named DBW(1), is obtained for $\lambda=\mu$ and $c_0<-2a$ and it results on the following warp factor \cite{degen3}
\begin{eqnarray}\label{warp_degen1}
&& e^{2A(y)}=n\left[ \frac{2a^2}{\left( \sqrt{c_{0}^{2}-4a^2}\right)
\cosh (2\mu a y)-c_{0}}\right]^{4a^2/9}  \nonumber \\
&& \times \exp \left\{ \frac{2a^2\left[ c_{0}^{2} - 4a^2-c_{0}\left(
\sqrt{c_{0}^{2}-4a^2}\right) \cosh (2a\mu y)\right] }{9\left[ \left( \sqrt{
c_{0}^{2}-4a^2}\right) \cosh (2a\mu y)-c_{0}\right] ^{2}}\right\} ,
\end{eqnarray}
where $n$ is chosen such that $e^{2A(0)}=1$.

The warp factor and the resulting potential $V(z)$ of the Schr\"{o}dinger-like Eq. \eqref{eq:schro} are plotted in Fig. \ref{fig:resd1} on the left side and in the center, respectively. As we can observe, the $c_0$ constant plays a role similar to the parameter $r$ that controls the thickness of the standard Bloch brane solution. However, the integration constant $c_0$, unlike the parameter $r$, is not present in Lagrangian density. When $c_0$ approximates to its critical value $-2a$ the minimum of the potential splits in two and we observe the appearance of a flat region on the warp factor. Another interesting feature is the formation of a two-kink solution in $\phi(y)$ when $c_0$ approximates to its critical value. Such feature was understood on the work \cite{degen2} as the formation of a double wall structure along the extra dimension.

From the $A(y)$ solution in Eq. \eqref{warp_degen1}, we compute the relative probability \eqref{rel_prob} again and the results are showed on the right side of the Fig. \ref{fig:resd1}. Sweeping the region where $m^2<V_{max}$, the function $N(m)$ show us resonance peaks at $m=0$. Such resonances correspond to the graviton zero modes. Varying $c_0$ and approaching the critical value, the resonances are not significantly changed.

New results are obtained when we consider another degenerate Bloch wall, named DBW(2). Such solution is obtained when $\lambda=4\mu$ and $c_0<1/16a^2$.  The corresponding warp factor is
\begin{eqnarray}
&& e^{2A(r)}=n\left[ \frac{2a}{\sqrt{\left( \sqrt{1-16c_{0}a^2}\right)
\cosh (4\mu ay)+1}}\right] ^{\frac{16a^{2}}{9}}  \nonumber \\
&& \times \exp{\left\{-\frac{4a^{2}\left[1+8a^{2}c_{0}+\left(
\sqrt{1-16c_{0}a^2}\right) \cosh (4\mu ay)\right]}{9 \left[ 1+\left( \sqrt{
1-16c_{0}a^2}\right) \cosh (4\mu a y)\right] ^{2}} \right\}}.
\label{2.93}
\end{eqnarray}
In this case, we also find a double kink profile to $\phi(y)$ when the integration constant approaches to the critical value.

\begin{figure}[!htb] 
      \centering
       \begin{minipage}[b]{0.95 \linewidth}
           \includegraphics[width=\linewidth]{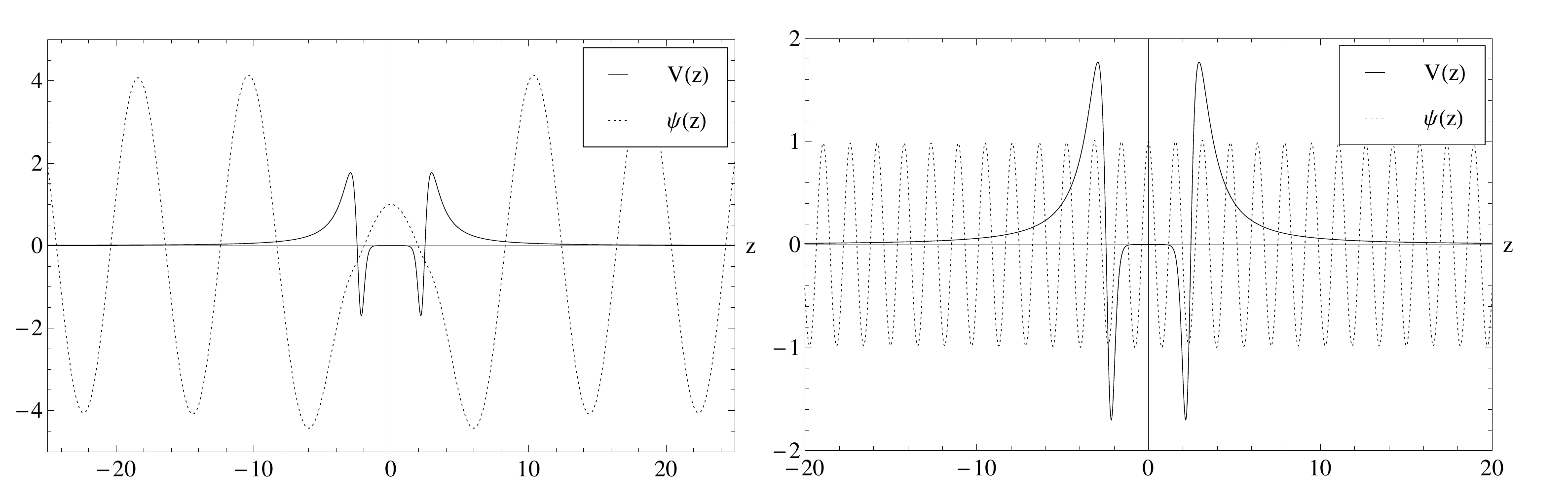}\\
           \caption{Plots of the potential $V(z)$ and the solution $\psi(z)$ for $m=0.8$ (\textit{left}) and $m=4$ (\textit{right}). We have used $a=\mu=1$ and $c_0=1/16.000001$ in the two graphics.}
       \label{fig:pot}
       \end{minipage}
   \end{figure}

\begin{figure}[!htb] 
      \centering
       \begin{minipage}[b]{0.6 \linewidth}
           \includegraphics[width=\linewidth]{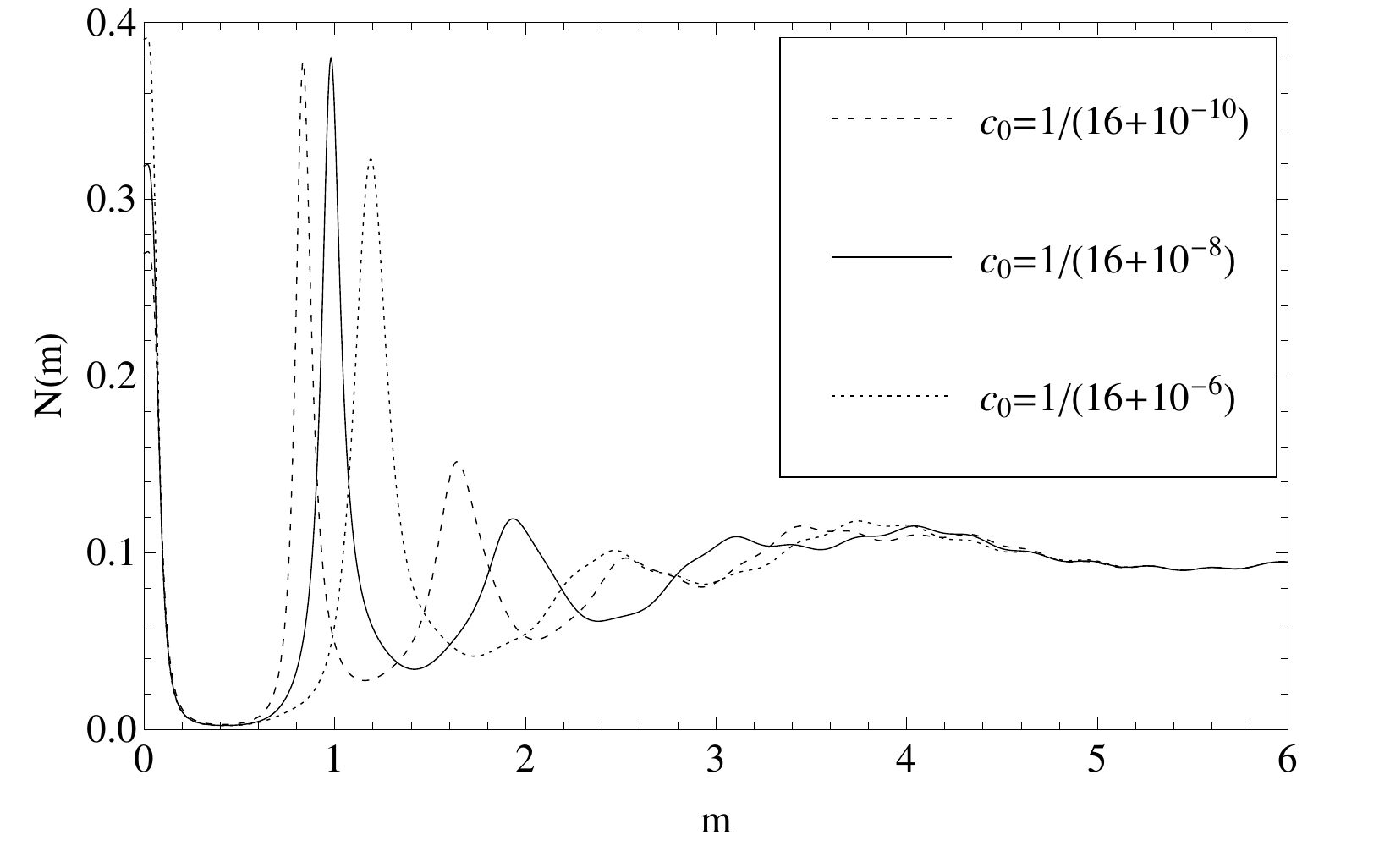}\\
           \caption{Plots of function $N(z)$ for DBW(2) solution. We have used  $a=\mu =1$ and $c_0<1/16$.}
       \label{fig:resd2}
       \end{minipage}
   \end{figure}

The potential and some solutions to $\psi(z)$ are plotted in Fig. \ref{fig:pot} for $a=\mu=1$ and $c_0=1/(16+10^{-6})$. When $m^2>V_{max}$ the solutions oscillate rapidly along the extra dimension, what is observed on the right side of the Fig. \ref{fig:pot}. However, as noted in the plot on the left,  for $m^2\leq V_{max}$ the amplitude of the wave solutions are suppressed in the region between the maxima of the potential and the modes  could potentially exhibit a resonance structure \cite{gremm, csaki1, csaki2}. For $c_0=1/(16+10^{-6})$ we detect two resonances, as showed on the Fig. \ref{fig:resd2}. Beyond the peak corresponding to the zero mode ($m=0$) we find another resonance at $m=1.1894$. Such result can be understood as the existence of graviton massive modes highly coupled to the brane.

When the constant $c_0$ approximates its critical value, the peak at $m=0$ is kept and the number of resonances with $m\neq0$ increases. The presence of resonances at $m=0$ for the cases considered shows the consistency of the relative probability method applied. Indeed, the Bloch branes support graviton zero modes localized and therefore there should be resonances corresponding to the zero modes.

The number of resonances increases when $c_0$ tends to $1/16$ and their lifetimes are increased as well. With $c_0=1/(16+10^{-10})$, for example, we have two peaks of probability beyond the peak at $m=0$. Such massive modes have high amplitudes at $z=0$ and should interact with the brane.

\section{Conclusions}\label{sec.conc}

We have considered a thick brane generated by two scalar fields in which we search for graviton resonances. We have reviewed the brane setup known as Bloch brane  in the Sec. 2. We know from the literature \cite{dionisio} that the real parameter $r$ controls the Bloch brane thickness, which in the interval $0.17>r>0$ produces a splitting on the defect. We have analyzed such property in terms of the curvature invariant. The results show that when we increase the brane thickness we observe a splitting effect on the maximum of the curvature invariant and the raising of a flat region at y=0.

From a metric perturbation, we find a Schr\"{o}dinger-like equation of motion. Since the previous works \cite{dionisio,degen1,degen2,degen3} have studied the zero mode cases, we started the analysis to identify resonances comparing two probability methods from $m\neq0$ Schr\"{o}edinger-like equations. The outcomes indicated that the direct probability method $P(m)$ has shown a resonance correctly. However, in the $N(m)$ function of relative probability the resonance peak was more pronounced indicating this process as more effective for detecting possible narrow resonances. We tested the efficacy  of the relative probability method to identify resonances by changing the integration range $z_b$. In this case, the resonance position was not changed.

The relative probability function $N(m)$ in Eq. \eqref{rel_prob} provide us a resonance at $m=0$. This result is in agreement with previous results which demonstrated the existence of a zero mode located \cite{dionisio}. On the other hand, looking for the effects of the internal structure on the resonance we have noted that when we reduce $r$, which makes the brane thicker, the width of the resonance is increased. Thus reducing $r$ will decreases the resonance lifetime. The same effect was observed at the zero mode which corresponds to the resonances, as shown in Fig.  \ref{fig:gravbloch3}. Furthermore, reducing $r$ and making the brane thicker causes a splitting of the defect that delocalizes the zero modes. Similar feature was also observed in the Ref. \cite{fase} as a phase transition in thick branes.

The existence of a resonance at zero energy is related to the existence of scales on which 4D gravity is recovered. On this way, when the resonances acquire large lifetimes we can restore the RS scenario \cite{RS}. Otherwise, when the width of the resonance increases too much, we will no longer have a region of 4D gravity \cite{csaki2,metastable}. The results obtained show that when we increase the thickness of the Bloch brane (reducing $r$), we reduce the lifetime of the graviton resonance. In this case and we can not guarantee the existence of a long-distance scale where the Newtonian potential is valid. In the same way, if we made the brane thicker, the localization of the zero modes could be jeopardized. We conclude that in the limit when $r \rightarrow 1/2$ the Bloch brane supports a region where gravity is effectively 4D. However, when $r\rightarrow 0$ the brane becomes very thick and this result is no more valid.

We have extended the investigation to the degenerate Bloch brane solutions. In the first degenerate case examined we have a critical constant $c_0$  that plays a role similar to the parameter $r$ in Sec. 3.  When $c_0$ approximates to its critical value $2a$ the brane becomes thick and the minimum of the potential splits in two.  The analysis shows resonances at $m=0$ corresponding to the graviton zero modes. They should be understood the same manner as those found in the basic Bloch brane setup.

We find interesting and new issues in the second class of degenerate solutions. We observe massive graviton resonances in addition to those at $m=0$. The appearance of resonant modes happens when the degeneracy parameter approaches its critical value. The number of resonances and their respective lifetimes increases when $c_0$ tends to $1/16$.  Another striking point is that when the brane is thicker ($c_0 \rightarrow 1/16$) the resonance lifetimes increases. Such behaviour is opposite to that find in the first Bloch brane scenario examined. The presence of those structures in the massive spectrum is related to the existence of massive (quasilocalized) modes highly coupled to the brane. We find massive resonances to the graviton field in another context of thick branes in Ref. \cite{nosso2}.

\textit{Note added}. -During the process of revision of this work  we find another study \cite{novochineses}  of resonances in two-field thick brane scenarios. The authors also noticed the presence of graviton resonances.

\section*{Acknowledgments}
The authors thank the Funda\c{c}\~{a}o Cearense de apoio ao Desenvolvimento
Cient\'{\i}fico e Tecnol\'{o}gico (FUNCAP), the Coordena\c{c}\~{a}o de Aperfei\c{c}oamento de Pessoal de N\' ivel Superior (CAPES), and the Conselho Nacional de Desenvolvimento Cient\' ifico e Tecnol\' ogico (CNPq) for financial support.


\begin{thebibliography}{99}

\bibitem{rub} V.A. Rubakov and M.E. Shaposhnikov, Phys. Lett. B 125 (1983) 136.
\bibitem{vis} M. Visser, Phys.Lett. B 159 (1985) 22.

\bibitem{wall} P. Coullet, J. Lega, B. Houchmandzadeh, and J. Lajzerowicz, Phys. Rev. Lett. 65 (1990) 1352.

\bibitem{dionisio} D. Bazeia and A. R. Gomes, J. High Energy Phys.
0405 (2004) 012.
\bibitem{aplications} D. Bazeia, J. Menezes, and R. Menezes, Phys.
Rev. Lett. \textbf{91} 241601 (2003).

\bibitem{brane} D. Bazeia, C. Furtado and A.R. Gomes, J. Cosmol.
Astropart. Phys. 0402 (2004) 002.
\bibitem{degen1} A. de Souza Dutra, Phys.Lett. B \textbf{626}, 249 (2005).

\bibitem{degen2} A. de Souza Dutra , A.C.Amaro de Faria, Jr. and M. Hott,  Phys. Rev. D \textbf{78}, 043526 (2008).

\bibitem{degen3} R. A. C. Correa, A. de Souza Dutra and M.B. Hott, Class. Quant. Grav. \textbf{28}, 155012 (2011).
\bibitem{wetting1} R. Lipowsky, Phys. Rev. Lett. \textbf{49}, 1575 (1982).

\bibitem{wetting2} R. Lipowsky, Phys. Rev. Lett. \textbf{52}, 1429 (1984).

\bibitem{fase} A. Campos, Phys. Rev. Lett \textbf{88}, (2002) 141602.


\bibitem{castro} L.B. Castro, Phys. Rev. D \textbf{83} 045002 (2011).

\bibitem{ca} C. A. S. Almeida, M. M. Ferreira, Jr., A. R. Gomes and
R. Casana, Phys. Rev. D \textbf{79} 125022 (2009).

\bibitem{chineses0}  Zhen-Hua Zhao, Yu-Xiao Liu and Hai-Tao Li, Class. Quant. Grav. \textbf{27} (2010) 185001.

\bibitem{nosso6} W. T. Cruz, Aristeu R. P. Lima and C. A. S. Almeida,
Phys. Rev. D \textbf{87} 045018 (2013).

\bibitem{nosso7} W. T. Cruz, R. V. Maluf and C. A. S. Almeida, Eur. Phys. J. C {\bf 73}, 2523 (2013).


\bibitem{gremm} M. Gremm, Phys. Lett. B \textbf{478} 434 (2000).



\bibitem{nosso} M. O. Tahim, W. T. Cruz and C. A. S. Almeida, Phys.
Rev. D \textbf{79} 085022 (2009).

\bibitem{csaki1} C. Csaki, J. Erlich, T. J. Hollowood and Y. Shirman,
Nucl. Phys. B \textbf{581} 309 (2000).

\bibitem{chineses1} Yu-Xiao Liu, Hai-Tao Li, Zhen-Hua Zhao, Jing-Xin
Li and Ji-Rong Ren, J. High Energy Phys. 10 (2009) 091.

\bibitem{chineses2} Yu-Xiao Liu, Chun-E Fu, Heng Guo and Hai-Tao
Li, Phys. Rev. D \textbf{85} 084023 (2012).

\bibitem{chineses3} Yu-Xiao Liu, Jie Yang, Zhen-Hua Zhao, Chun-E
Fu and Yi-Shi Duan, Phys. Rev. D \textbf{80} 065019 (2009).

\bibitem{chineses4} Yu-Xiao Liu, Chun-E Fu, Li Zhao, Yi-Shi Duan,
Phys. Rev. D \textbf{80} 065020 (2009).

\bibitem{chineses5} Yun-Zhi Du, Li Zhao, Yi Zhong, Chun-E Fu and
Heng Guo, \textit{Resonances of Kalb-Ramond field on symmetric and asymmetric thick branes}, arXiv:hep-th/13013204.

\bibitem{csaki2} C. Csaki, J. Erlich, T. J. Hollowood and Y. Shirman,
Phys. Rev. Lett. \textbf{84} 5932 (2000).

\bibitem{metastable} G. Dvali, G. Gabadadze, M. Porrati, Phys. Lett. B \textbf{484} (2000) 112.

\bibitem{bazeia1}D. Bazeia, M.J. dos Santos and R.F. Ribeiro, Phys.
Lett. A \textbf{208} 84 (1995).

\bibitem{bazeia2}D. Bazeia, H. Boschi-Filho, and F.A. Brito, J. High
Energy Phys. 9904 (1999) 028.

\bibitem{bazeia3}D. Bazeia and F.A. Brito, Phys. Rev. Lett. \textbf{84},
1094 (2000).

\bibitem{bazeia4} D. Bazeia, J.R. Nascimento, R.F. Ribeiro and D.
Toledo, J. Phys. A \textbf{30}, 8157 (1997).

\bibitem{shif} M.A. Shifman and M.B. Voloshin, Phys. Rev. D \textbf{57},
2590 (1998).

\bibitem{alonso} A. Alonso Izquierdo, M.A. Gonzalez Lion and J. Mateos
Guilarte, Phys. Rev. D \textbf{65}, 085012 (2002).

\bibitem{kehagias} A. Kehagias and K. Tamvakis, Phys. Lett. B \textbf{504},
38 (2001).

\bibitem{de} O. Dewolfe, D.Z. Freedman, S.S. Gubser and A. Karch,
Phys. Rev. D \textbf{62}, 046008 (2000).

\bibitem{nosso1} W.T. Cruz, M.O. Tahim and C.A.S. Almeida, Europhys.
Lett. \textbf{88}, 41001 (2009).

\bibitem{lifetime} R. Gregory, V. A. Rubakov and S. M. Sibiryakov,
Phys. Rev. Lett. \textbf{84}, 5928 (2000).

\bibitem{RS} L. Randall and R. Sundrum, Phys. Rev. Lett. \textbf{83}
3370 (1999); \textbf{83}, 4690 (1999).



\bibitem{nosso2} W. T. Cruz , A. R. Gomes and C. A. S. Almeida, Europhys. Lett. \textbf{96}, 31001 (2011).

\bibitem{novochineses} Qun-Ying Xie, Jie Yang and Li Zhao, Phys. Rev. D \textbf{88}, 105014 (2013).



\end{thebibliography}
\end{document}